\documentclass[12pt]{iopart}
\usepackage{amssymb, hyperref, graphicx}
\usepackage{iopams}  
\usepackage{caption}
\begin{document}

\title[Blurred path-spin entanglement]{Blurred path-spin entanglement in Stern-Gerlach apparatus:\\
interplay between magnetic inhomogeneity and Larmor precession}

\author{Nirupam Dutta $^1$, Ansuman Dey $^2$}
\address{$^1$Theoretical Physics Division, Variable Energy Cyclotron Centre, 1/AF Bidhan Nagar, Kolkata 700~064, India}\ead{nirupamdu@gmail.com}
\address{$^2$S. N. Bose National Centre For basic Sciences, Saltlake, Kolkata-700098, India}\ead{ansuman@bose.res.in}

\vspace{10pt}

\begin{abstract}
We argue that the non-adiabatic evolution of spin states in Stern-Gerlach apparatus can blur the 
manifestation of path spin entanglement. This fact questions the usual practice of spin measurement even in a formally and operationally ideal situation. Considering azimuthal inhomogeneity, we have identified through quantitative calculation the specific reason behind the breakdown of adiabatic evolution to be the spatial inhomogeneity of applied magnetic field.  The angle $\theta$ between the $z$ component of magnetic moment of a particle and direction of applied magnetic field is also an important factor in determining the category of evolution of spin states.  Adiabaticity always can be restored by choosing a sufficiently small value of $\theta$.  
\end{abstract}

\maketitle

\section{Introduction}
The Stern-Gerlach experiment (SGE)\cite{gerlach1922} plays a fundamental role in quantum mechanics because of its conceptual relevance. SGE first witnessed the quantum nature of intrinsic spin of a particle. Besides realizing the existence of spin angular momentum and its space quantization, the experiment became a formal ground to study decoherence and quantum non locality like entanglement \cite{reinisch1999, englert1988,einstein1922}. Different aspects of SGE are in general quite complicated and are still being studied by present day researchers  \cite{hsu2011} . In the conventional description of the experiment, a beam of particles is passed through a Stern-Gerlach(SG) apparatus which consists of a magnet providing an inhomogeneous magnetic field, in a certain direction(z, say). When a screen is put at a distance from where the particles are emerging from the SG apparatus, two distinct peaks are observed in positive and negative $z$ directions corresponding to two different spin values along $z$ axis. This clearly demonstrates a path spin entanglement. More precisely, the eigenstates of $S_{z}$ of a charge neutral particle entering the SG apparatus are entangled with its spatial degrees of freedom. The observed peaks on the screen always lead to the detection of up spin in the upward ($z>0$) and down spin in the downward ($z<0$) direction. This statement is not only true for the measurement of spin along $z$ axis, but is also valid when the spin is measured along any arbitrary direction $\hat{u}$ along which the magnetic field is applied. The magnetic field spatially separates two different spin states $|+\rangle_{u}$ and $|-\rangle_{u}$ and they evolve being coupled with two different parts of the wave function. The final state becomes $|\psi \rangle =c_+\phi_{+}({\bf x}) \otimes |+\rangle_{u} + c_{-}\phi_{-}({\bf x})\otimes|-\rangle_{u}$ with $c_{+}$, $c_{-}$ complex constants. Here, $\phi_{+}({\bf x})$, $\phi_{-}({\bf x})$ are the spatial parts in upward and downward directions of $u=0$ plane while $|+\rangle_{u}$, $|-\rangle_{u}$ are the eigenstates of $\hat{S_{u}}$. ${\bf x}$ denotes the spatial degrees of freedom of the particle. In a ideal SG experiment, the up (down) spin can only be found in upward (downward) direction. This is possible only when the deflected beams are well separated such that $\phi_{+}({\bf x})$ and $\phi_{-}({\bf x})$ are orthogonal inside the SG apparatus and the orthogonality is also preserved during the free evolution of the particles as they leave the apparatus and travel to the screen. Hence, the overlap function
\begin{equation}
I = \int_{-\infty}^{\infty} \phi_{+}(\bold{x}) \phi_{-}(\bold{x})d^3 \bold{x}=0. \label{orthogonality}
\end{equation}
This is referred to as a formally ideal situation \cite{home2007}. An additional requirement of operational idealness \cite{home2007} may also be invoked where 
\begin{eqnarray}
\fl  E=\int_{x=-\infty}^{\infty} \int_{y=-\infty}^{\infty} \int_{z=0}^\infty |\phi_{-}(\bold{x})|^2 dx dy dz=\int_{x=-\infty}^{\infty} \int_{y=-\infty}^{\infty} \int_{z=-\infty}^0 |\phi_+(\bold{x})|^2 dxdydz = 0~.\label{opid}
\end{eqnarray}
Any exception to the above conditions is known to lead to a non ideal outcome \footnote{Note that the definitions of both formal and operational idealness concern only the spatial parts of the state of the particle}. However, we shall argue that these criteria, though necessary, are not sufficient to ensure an ideal outcome. Till date, another important assumption has been implicitly incorporated in the analysis of this experiment -- it is the adiabatic evolution of spin states. By adiabatic evolution \cite{griffithsqm,messiahqm} we mean that the $|+\rangle$ ($|-\rangle$) state which is coupled to $\phi_+$ ($\phi_-$) remains in the same state during its travel through the apparatus. But it can not be true in every situation, especially in a weak magnetic environment. This non adiabatic evolution of spin states will be shown to lead to the emergence of non idealness even in a `formally and operationally ideal' case ($I =0,E=0$).

Though most studies on conceptual and experimental aspects of SGE have been conducted within the confines of adiabatic regime, there also are instances in literature \cite{bliokh2007,papanicolaou1988} where the concept of adiabaticity has been addressed for explicitly time varying magnetic fields. However, the spatial variation of static inhomogeneous magnetic field can also play a crucial role in the measurement of spin through path spin entanglement. Spatial inhomogeneity of the magnetic field lends an implicit time dependence to the interaction Hamiltonian which becomes explicit in the rest frame of the particle. A rapid spatial variation of interaction Hamiltonian in that case may cause a non adiabatic evolution for spin states. This gives rise to the possibility of detecting both up and down spins in either direction. Hence, one cannot infer a definite spin of the particle by observing its deflection although the split in the distribution pattern on the screen is still there (due to the force on the particles caused by the magnetic field). 
  In other words, though the observed state on the screen is, in general, a path spin entangled one, a specific path does not necessarily correspond to a particular spin. This, we call {\it blurred path spin entanglement}. In fact, under certain conditions, instead of down spin the up spin can couple with $\phi_{-}$ and vice versa. This is absolutely opposite to the conventional wisdom and obtaining such a results especially from the analysis of a formally and operationally ideal SG experiment is a highly non trivial issue.

In order to demonstrate our finding, we have organised the article in the following way. After the introduction, we have discussed the relevance of adiabatic and non-adiabatic evolution of spin$\frac{1}{2}$ particles in the context of SG experiments. Section $3$ is devoted to examining the time evolution of spin states by using Schroedinger equation. In this context, we also have pointed out the parameters which decide the category (adiabatic or non-adiabatic) of the time evolution. This result is carried over to section $4$ for the interpretation of blurred path spin entanglement and its non trivial consequences. Finally, to summarise, a concise discussion is presented at the end of the article.

\section{Adiabaticity and path spin entanglement}
Suppose a bunch of charge-neutral spin$\frac{1}{2}$ particles, prepared in a superposed state of up $|+\rangle _{u}$ and down $|-\rangle_{u}$ spins are collimated towards an SG apparatus. The state of a particle before it enters the device can be expressed as,
\begin{equation}
|\psi_{i}\rangle = \big(c_{+} |+\rangle _{u}+ c_{-} |-\rangle _{u} \big)\otimes \phi_{i}(\bold{x}),
\end{equation}
where $\phi_{i}(\bold{x})$ is the spatial part of the state. The moment the particle enters the inhomogeneous magnetic field along some direction $\hat{u}$, the state becomes path spin entangled due to its interaction with the magnetic field and can be written as,
\begin{equation}
|\psi\rangle =  c_{+} |+\rangle _{u}\otimes \phi_{+}({\bf x})+ c_{-} |-\rangle _{u} \otimes  \phi _{-}({\bf x}).\label{ent}
\end{equation}
As a result, a part of the particle beam will start to propagate along the upward direction and the rest will get directed downward. This fact is evident from the above equation which has already been derived many times in this context. In conventional treatments inspired by the original Stern-Gerlach paper \cite{gerlach1922}, the inhomogeneous magnetic field is assumed to be directed along a fixed direction (say $z$). However, this does not make the field divergence free, thus contradicting Maxwell's equations \cite{platt1992,aharonov1988}. As a way out, the authors in \cite{platt1992,oliveira2006} introduced a two component magnetic field and adopted a time averaged description of Pauli equation thereby obtaining an effective magnetic field along a fixed direction. This assumption is based on the consideration of very strong magnetic fields \cite{platt1992,alstrom1982} observed over a time scale much larger than the characteristic time scale of the spin half system. Obviously, it is not impossible to design such a circumstance with sufficient control over laboratory conditions. But the same cannot be said of general practical scenarios. The Stern-Gerlach model is widely used to study path spin entanglement in diverse magnetic environments where this assumption may not always hold. Especially for a weak magnetic field, the above mentioned treatment is far from warranted. In a more general situation, the inhomogeneous magnetic field inside the apparatus changes spatially not only in magnitude but also in its direction. One can always think of a collection of tiny SG apparatus each providing magnetic fields in different directions so that they together mimic the effect of a single apparatus having a magnetic field varying along the length.

Due to such inhomogeneity, the particle will experience a varying magnetic field during its flight from one space point to another. So, it will see a magnetic field which has an implicit time dependence. We would like to examine the situation for upward and downward beams separately from the rest frames of the corresponding particles. Here it is worth mentioning that the implicit time dependence of the magnetic field $B$ becomes explicit in the rest frame of the particle. Let us call the interaction Hamiltonian in the rest frame of the particle $H_{part}(t)$. It is clear that the interaction Hamiltonian $H_{part}(t)$ can not affect the spatial parts $\phi_+$ and $\phi_-$ as it no longer contains any spatial degree of freedom. Therefore, we need to study the evolution of up $|+\rangle$ and $|-\rangle$ spin states only. Before proceeding, below we justify and explain our scheme of working in the particle rest frame.

Obviously, the particle rest frame is not ideally inertial. Note that in an SG apparatus, interaction with a weak magnetic field can cause only a small change $\Delta p $ in the momentum of an outgoing particle. This small change is negligible compared to the momentum $p$ of a high energy incoming beam. Thus, taking the particle frame to be inertial is practically a nice approximation in this context, unlike in a strong magnetic environment where $\frac{\Delta p}{P}$ is not likely to be much less than unity. But an exception occurs even in the latter case when a particle has a de Broglie wavelength $\lambda$ comparable to the length of the apparatus. This is because the particle does not experience a noticeable change in its momentum inside the apparatus as the dimension of the SG apparatus is not much bigger than the size of the wave packet. Except for these two special cases, one must take into account non-inertial modifications of the Hamiltonian while working in the particle rest frame. In the present article we restrict ourselves to the two special simple cases only.

Suppose at some instant, the direction of $\vec{B}_{part}(\tau)$, the magnetic field seen from the particle rest frame, inside the apparatus is along $\hat{u}(\tau)$. $H_{part}(\tau)$ and $S_{u(\tau)}$ have common set of eigenvectors $|+\rangle_{u(\tau)}$ and $|-\rangle_{u(\tau)}$ as,
\begin{equation}
H_{part}(\tau)= -\gamma B_{part}(\tau) S_{u(\tau)},\label{gyro}
\end{equation}
where $\gamma$ is the gyromagnetic ratio of the particle.
 Now, the task is to understand whether the spin states evolve adiabatically or not under the action of such a time dependent Hamiltonian. If the Hamiltonian changes sufficiently slowly with time such that 
\begin{equation}
\fl \Bigg |\frac{_{u(\tau)}\langle \pm|\frac{d}{d\tau}|\mp\rangle_{u(\tau)} }{E_{\pm}-E_{\mp}}\Bigg |=\bigg |\frac{_{u(\tau)}\langle \pm|\dot{H}_{part}|\mp \rangle_{u(\tau)}}{\Big(E_{\pm}-E_{\mp}\Big)^{2}}\bigg|
=\bigg |\frac{_{u(\tau)}\langle \pm|\vec{\mu}.\dot{\vec{B}}_{part}|\mp \rangle_{u(\tau)}}{\Big(E_{\pm}-E_{\mp}\Big)^{2}}\bigg|<<1~, \label{adiabatic}
\end{equation} 
then up (down) spin state will evolve to up(down) spin state of the corresponding instantaneous Hamiltonian $H_{part}(\tau)$ \cite{born1928,aharonov1987,amin2009} but with some geometrical (Berry phase)  and dynamical phase factors \cite{berry1984}. $|\mp \rangle_{u(\tau)}$ is the eigenstate corresponding to the energy eigenvalue $E_{\mp}$ and $\vec{\mu}$ is the magnetic moment of the particle. This implies that the upward (downward) beam always contains the up (down) spin only. In those situations for which the above condition is not satisfied, the evolved state in each direction becomes a superposition of up and down spins. Therefore, in the upward direction we will have finite transition probability to $|-\rangle_{u(\tau)}$ from $|+\rangle_{u(0)}$ and similarly to the state $|+\rangle_{u(\tau)}$ from$|-\rangle_{u(0)}$ in the downward direction. As a result, one will find both up and down spins in either direction even though a formally and operationally ideal situation has been maintained by keeping $I=0=E$.  This reveals the so far overlooked, salient role played by the adiabatic approximation in an SG experiment.

\section{Interplay between Larmor precession and azimuthal inhomogeneity of magnetic field}

In this section we explicitly examine the evolution of spin states in an inhomogeneous magnetic field keeping aside the assumptions made in earlier literature \cite{platt1992,potel2005}. We avoid any time averaged description of Pauli equation so that our analysis can be applied to a more general scenario. In the rest frame of the particle, the magnetic field $B_{part}$ appears to change with time. So, by considering an azimuthally inhomogeneous magnetic field interacting with the magnetic dipole moment $\vec{\mu}$, the explicit time dependence of the Hamiltonian can be expressed as,
\begin{figure}[t]
\centering
\includegraphics[scale=0.48]{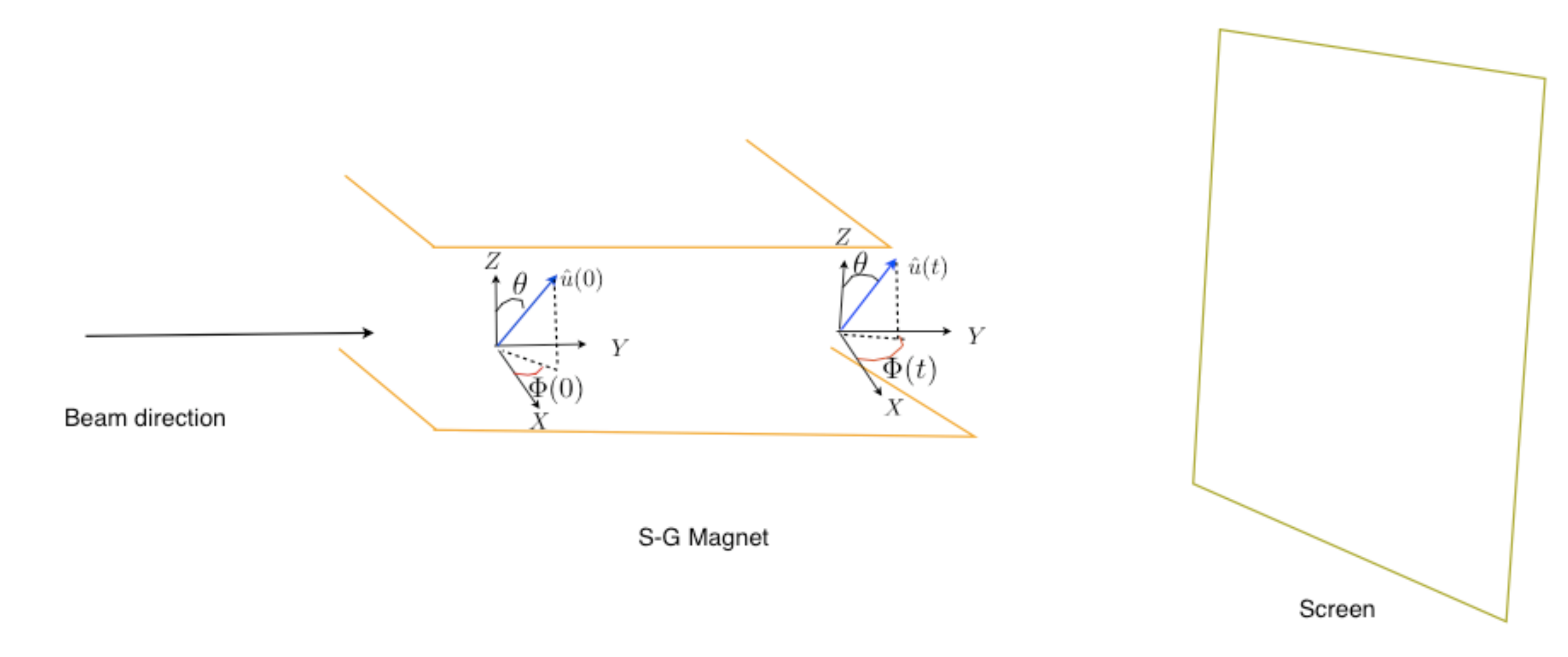}
\caption{A schematic diagram which shows how the direction $\hat{u}$ of the magnetic field $\vec{B}_{part}$ appears to be changing with time from the rest frame of the particle as it moves through the SG apparatus.}\label{dia}
\end{figure}

\begin{eqnarray}
H_{part}(t) 
&=-\vec{\mu}.\vec{B}_{part}(\tau)=-\big(\vec{\mu}.\hat{u}(\tau)\big)B_{part}=\omega_{0}\hat{S}_{u(\tau)} \nonumber \\
&=\frac{\omega_{0} }{2}\Big(\sigma_{x}\sin \theta \cos\Phi(\tau) +  \sigma_{y}\sin\theta \sin\Phi(\tau) +\sigma_{z}\cos\theta \Big) \label{Hamiltonian}.
\end{eqnarray} 
Here, $\hat{u}(\tau)$ is the direction of the magnetic field $\vec{B}_{part}$ and the azimuthal inhomogeneity makes the Hamiltonian a time dependent quantity through $\Phi(\tau)$. The  Larmor frequency $\omega_{0}$ is taken to be a constant quantity by assuming that the magnitude of the magnetic field is not changing appreciably inside the apparatus. $\sigma_{x}$, $\sigma_{y}$, $\sigma_{z}$ are the Pauli spin matrices and $\theta$ is the angle which the magnetic field makes with $z$ axis. 
We have already mentioned in the last section that after entering SG apparatus, the particle beam gets split into two different parts, expressed through a path spin entangled state,
 \begin{equation}
|\Psi(0)\rangle=c_+|+\rangle_{u(0)} \otimes\phi_{+}({\bf x},0)+ c_-|-\rangle_{u(0)}\otimes \phi_{-}({\bf x},0). \label{initial}
\end{equation}
 We denote by $\tau=0$ the instant when the particle just becomes path spin entangled inside the apparatus. Here $\phi_{+}(\bold{x},0)$ and $\phi_{-}(\bold{x},0)$ are the up and down components of the spatial wave function at $\tau=0$ with respect to the plane whose normal makes an angle $\theta$ with $z$ axis. For $\theta=0$, this plane is nothing but the $z=0$ plane and we get the usual text book set up. Now, we analyse both the beams separately from the corresponding rest frame of the particles. Let us say that the particle  leaves the apparatus at some instant $\tau=t$ and then moves freely up to the screen. A schematic diagram is presented in fig. \ref{dia} to demonstrate the set up under consideration. So, the final state of the particle we are interested in is $|\Psi(t)\rangle$.
 
Whether the evolution of the spin states in both upward and downward directions is adiabatic or not is determined from eq.\ref{adiabatic} by making use of instantaneous energy eigenstates of $H_{part}(\tau)$,
\begin{eqnarray}
|-\rangle_{u(\tau)} =
\left(\begin{array}{c}
e^{-i \frac{\Phi(\tau)}{2}} \sin{\frac{\theta}{2}} \\
-e^{i\frac{\Phi(\tau)}{2}} \cos{\frac{\theta}{2}}
\end{array} \right)
\nonumber \\
|+\rangle_{u(\tau)} =
\left( \begin{array}{c} 
e^{-i\frac{\Phi(\tau)}{2}}\cos \frac{\theta}{2}\\
e^{i\frac{\Phi(\tau)}{2}}\sin \frac{\theta}{2}
\end{array}\right)
\label{inst_phi}
\end{eqnarray}
corresponding to the energy eigenvalues $E_{-}$ and $E_{+}$ respectively. Henceforth, we will denote $u(\tau)$ by the slightly more compact notation $u_{\tau}$.
So, the ratio
\begin{eqnarray}
\Bigg |\frac{_{u_{\tau}}\langle -|\frac{d}{d\tau}|+\rangle_{u_{\tau}} }{E_{+}-E_{-}}\Bigg |=\frac{\dot{\Phi}}{2\omega_{0}} \sin \theta \label{adbc} .
\end{eqnarray}
The above equation shows that $\theta$, $\dot{\Phi}$ and $\omega_{0}$ are the determining factors for the adiabatic evolution of spin states. For non zero values of $\sin\theta$, $\omega_{0}$ should be much larger than $\dot{\Phi}$ in order to satisfy the adiabatic condition \ref{adiabatic} . It implies that the Larmor precession should be rapid enough to cope with the instantaneous direction of the magnetic field.

In order to know the time evolution of spin states, one needs to solve the Schr\"odinger equation for the time dependent Hamiltonian $H_{part}(\tau)$ with a given initial state. The evolved state at any instant $\tau$ could be expressed in terms of instantaneous basis vectors,
\begin{equation}
|\tau \rangle = a(\tau)|+\rangle_{u_{\tau}} + b(\tau)|-\rangle_{u_{\tau}}.
\end{equation}
Plugging this into Schr\"odinger equation we get the following set of linear differential equations,
\begin{eqnarray}
\dot{a}(\tau)+\frac{i}{2}\big[\omega_{0}-\dot{\Phi}(\tau)\cos\theta\big]a(\tau) -i\frac{\dot{\Phi}(\tau)}{2}b(\tau)\sin\theta = 0\nonumber\\
\dot{b}(\tau)-\frac{i}{2}\big[\omega_{0}-\dot{\Phi}(\tau)\cos\theta\big]b(\tau) -i\frac{\dot{\Phi}(\tau)}{2}a(\tau)\sin\theta =0 \label{gensoln}
\end{eqnarray}
For any quantitative prediction we need to know the time dependence of $\Phi(\tau)$. To proceed further, we present a solution by considering  terms up to 1st order in the Taylor series expansion of $\Phi(\tau)$. A general solution is always possible but we have considered this simple case to show the drastic role played by the azimuthal inhomogeneity.
\begin{equation}
\Phi(\tau)= \Phi_{0}+ \frac{d\Phi(\tau)}{d\tau}\big |_{\tau=0}\tau+ \Or(2)\label{Taylor}
\end{equation}
This is actually equivalent to assuming a small change in velocity of the particle as it travels through the apparatus (mentioned in the previous section).
The instantaneous states look (by setting $\Phi_{0}=0$ without any loss of generality),
\begin{eqnarray}
|-\rangle_{u_{\tau}} =
\left(\begin{array}{c} 
e^{-i\frac{\omega \tau}{2}} \sin{\frac{\theta}{2}}\\
-e^{i\frac{\omega \tau}{2}} \cos{\frac{\theta}{2}}
\end{array}\right)
\nonumber \\
|+\rangle_{u_{\tau}} =
\left(\begin{array}{c} 
e^{-i\frac{\omega \tau}{2}} \cos \frac{\theta}{2}\\
e^{i\frac{\omega \tau}{2}}\sin \frac{\theta}{2}
\end{array}\right). 
\label{inststate}
\end{eqnarray}
The adiabatic condition \ref{adiabatic} becomes $\frac{\omega}{\omega_{0}} \sin \theta <<1$, where $\omega = \frac{d\Phi(\tau)}{d\tau}\big |_{\tau=0}$.
The spin part $|+\rangle_{u_0}$ in \ref{initial} which is coupled to the spatial part $\phi_{+}$ evolves to the final state
\begin{equation}
|t\rangle_{+} = \alpha_{+}|+\rangle_{u_{t}} + \alpha_{-}|-\rangle_{u_{t}}\label{state2}.
\end{equation}
at the instant $\tau=t$.
The coefficients $\alpha_{+}$ and $\alpha_{-}$can be evaluated by solving the differential equation \ref{gensoln} with the condition $a(0)=1$, $b(0)=0$ and $a(t)=\alpha_{+}$, $b(t)=\alpha_{-}$.  They are
\begin{eqnarray}
\alpha_{+} = \cos \frac{\bar{\omega}t}{2} -i\frac{\omega_{0}-\omega\cos\theta}{\bar{\omega}}\sin\frac{\bar{\omega} t}{2} \nonumber\\
\alpha_{-}=   i\frac{\omega\sin \theta}{\bar{\omega}}\sin\frac{\bar{\omega}t}{2} .\label{coeff1}
\end{eqnarray}
$\bar{\omega}$ is a combination of $\omega$ and $\omega_{0}$:
\begin{equation}
\bar{\omega}= \sqrt{\omega^{2}_{0}+\omega^{2} -2\omega_{0}\omega \cos\theta}\label{omegabar}.
\end{equation}
So, we witness that the up spins in the upward direction will generate down spins through their non-adiabatic evolution.

Similarly, the evolved state corresponding to the initial state $|-\rangle_{u_0}$ which is coupled with $\phi_{-}$ is given by,
\begin{equation}
|t\rangle_{-} = \beta_{+}|+\rangle_{u_{t}} + \beta_{-}|-\rangle_{u_{t}},\label{state}
\end{equation}
The coefficients $\beta_{+}$ and $\beta_{-}$ are determined in the same way by inserting the conditions $a(0)=0$, $b(0)=1$ and $a(t)=\beta_{+}$, $b(t)=\beta_{-}$ in Schr\"odinger equation \cite{tong2010}:
\begin{eqnarray} 
\beta_{-} = \cos \frac{\bar{\omega}t}{2} +i\frac{\omega_{0}-\omega\cos\theta}{\bar{\omega}}\sin\frac{\bar{\omega}t}{2} \nonumber\\
 \beta_{+}=  i\frac{\omega\sin \theta}{\bar{\omega}}\sin\frac{\bar{\omega}t}{2}~.\label{eq:coeff2}
\end{eqnarray}
This again shows a finite transition probability to up spin state from the down spin in the downward direction. Therefore, we have up and down spins in both the directions. Hence, the usual perception of spin measurement through SG experiment can not be valid in general.
In principle, for any nonzero $\sin\theta$, one has to consider such non-adiabatic evolution of states whenever the quantity $\omega_{0}$ is not very large compared to $\omega$. 
\begin{figure}[!h]
\centering
\begin{minipage}{.45\linewidth}
  \includegraphics[width=\linewidth]{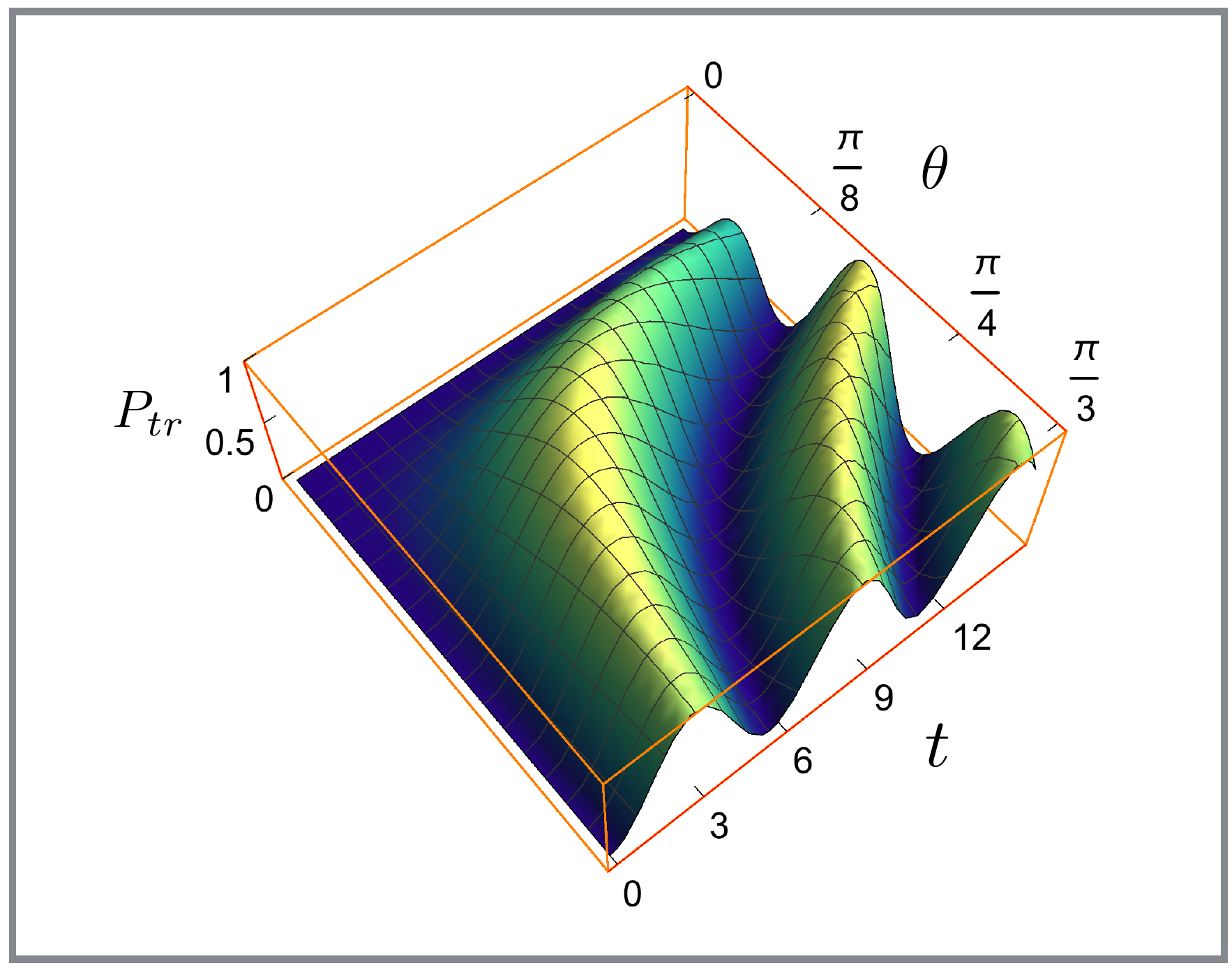}
  \caption{$P_{tr}$ as a function of time $t$ (in the unit of $\frac{1}{\bar{\omega}}$) and the angle $\theta$ for a chosen value of $\omega =1.2\omega_{0}$.} 
  \label{transition}
\end{minipage}
\hspace{.05\linewidth}
\begin{minipage}{.45\linewidth}
  \includegraphics[width=\linewidth]{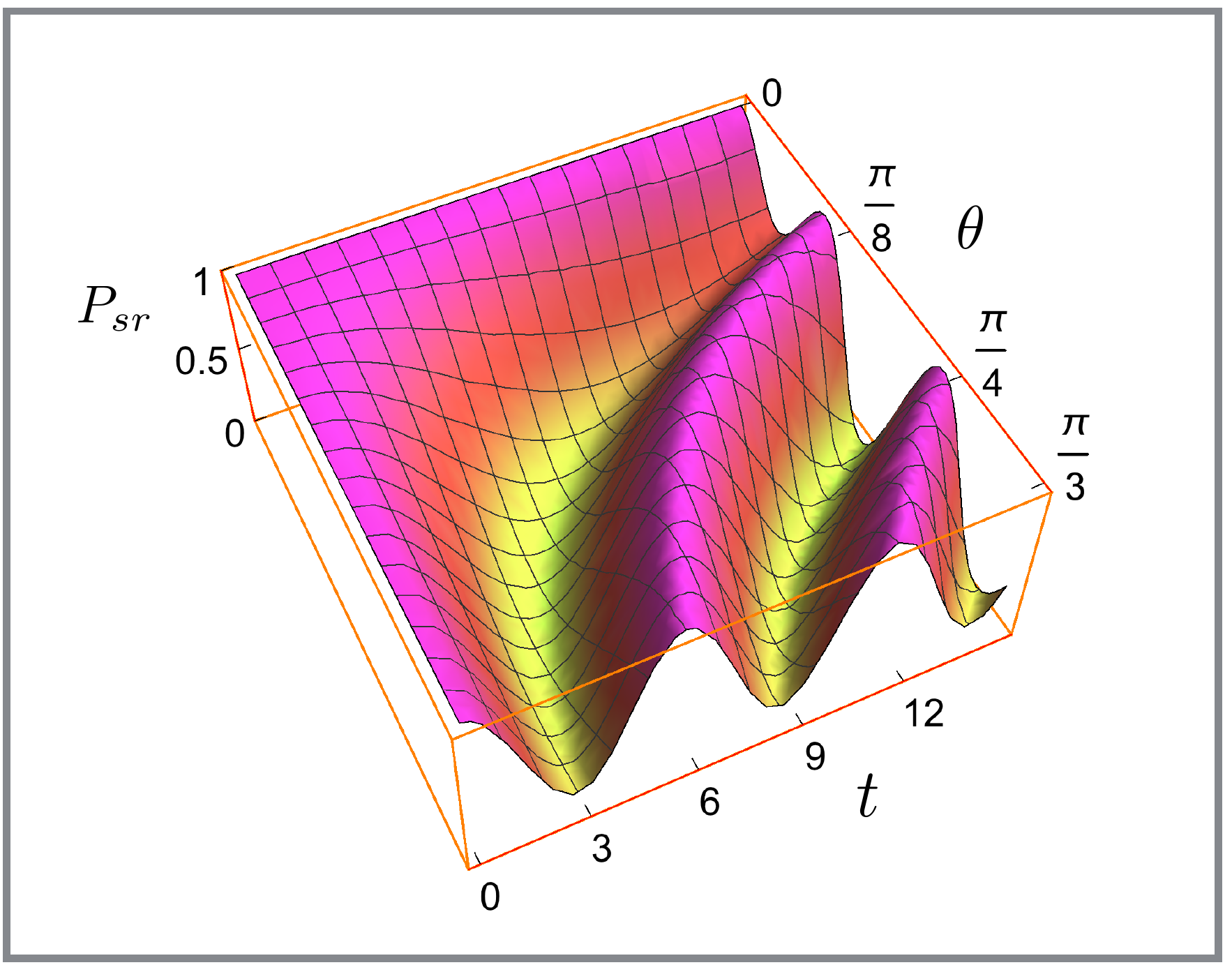}
  \caption{$P_{sr}$ as a function of time $t$ (in the unit of $\frac{1}{\bar{\omega}}$) and the angle $\theta$ for a chosen value of $\omega =1.2\omega_{0}$.} 
  \label{survival}
\end{minipage}
\end{figure}
Equation \ref{coeff1} and \ref{eq:coeff2} show that the probability of having down spin from up spin in the upward direction is $|\alpha_{-}|^{2}$ which is exactly equal to the transition probability $|\beta_{+}|^{2}$ from down spin to up spin in the downward direction. We call it $P_{tr}$ and present it graphically in Figure \ref{transition} as a function of $\theta$ and time $t$. The survival probability of up(down) spin states in the upward(downward) direction is equal to $|\alpha_{+}|^{2}$ ($|\beta_{-}|^{2}$). And it is clear that $|\alpha_{+}|^{2}=|\beta_{-}|^{2}$. Lets denote this probability as $P_{sr}$. Figure \ref{survival} shows its variation with $\theta$ and $t$. This is a very important consequence of the analysis. We see that the number of down (up) spins disappearing from the downward (upward) beam is exactly equal to the number of  down (up) spins appearing in the the upward (downward) beam. Therefore, as a whole the total spin is always conserved.

The final spin state which we observe on the screen is the one at the instant $t$ when the particle just leaves the apparatus. That instant has to be determined from the time taken by the particle to complete its travel through the SG apparatus. Therefore, the longitudinal dimension of the apparatus is important to explain the observed pattern on the screen.\\

It is also noticeable that the non adiabatically evolved spin states have a precession frequency which is different from the Larmor frequency $\omega_{0}$. The modification also depends on $\theta$ and $\omega$. This happens because the instantaneous spin eigenstates are now weighted by a periodic function ($\cos\bar{\omega}t$ or $\sin\bar{\omega}t$).

\section{Blurred path-spin entanglement}

In the previous section, we have derived the final spin states coupled to spatial up and down parts of $|\psi(0)\rangle$. Now, the final state $|\Psi(t)\rangle$ of the particle when it leaves the apparatus becomes,
\begin{eqnarray}
\fl |\Psi(t)\rangle &= c_{+}\big(\alpha_{+}|+\rangle_{u_{t}}+\alpha_{-}|-\rangle_{u_{t}}\big)\otimes \phi_{+}^{\prime}({\bf x^{\prime}},t)
 +  c_{-}\big(\beta_{+}|+\rangle_{u_{t}}+\beta_{-}|-\rangle_{u_{t}}\big)\otimes \phi_{-}^{\prime}({\bf x^{\prime}},t)\nonumber\\
\fl&= \big(d_{1}\phi_{+}^{\prime}(\bold{x^{\prime}},t)+d_{2}\phi_{-}^{\prime}({\bf x^{\prime}},t)\big)\otimes |+\rangle_{u_{t}} +\big(k_{1}\phi_{+}^{\prime}({\bf x^{\prime}},t)+k_{2}\phi_{-}^{\prime}({\bf x^{\prime}},t)\big)\otimes |-\rangle_{u_{t}} ~,\label{eq:soln}
\end{eqnarray}
where $\phi_{+}^{\prime}({\bf x^{\prime}},t)$ and $\phi_{-}^{\prime}({\bf x^{\prime}},t)$ are the spatial up and down components of the state as seen from the rest (primed) frame of the particle. We have mentioned earlier that the Hamiltonian $H_{part}(t)$ does not affect the spatial part of the state. The spatial up (down) part remains as the up (down) part  with respect to the plane whose normal is now $\hat{u}(t)$ even at the instant $\tau=t$ as long as the angle $\theta$ is constant. No wonder that the situation remains formally and operationally ideal if we start with $I=0, E=0$ at $\tau =0$. Equation \ref{eq:soln} shows that the final state is still an entangled one as the spatial and spin parts are not product separable. We  call it a path spin entangled state but a big difference from the conventional result is that the entangled state can no longer be used to
draw inference on the nature of particle spin from its path. In other words, the evolved state has  blurred the possibility of determining the spin of a particle from a knowledge of its path. Moreover, one can easily check that a breakdown of adiabaticity can not result in disentanglement of the initially path spin entangled state. Disentanglement occurs when $d_{1}=d_{2}=\frac{d}{\sqrt{2}}$ and $k_{1}=k_{2}=\frac{k}{\sqrt{2}}$ which can not be achieved for any value of $\theta$ or $t$.

However, at the instant $t=\frac{(2n+1)\pi}{\bar{\omega}}$, $n$ being any positive integer, when $\cos\theta = \frac{\omega_{0}}{\omega}$,
\begin{equation*}
\alpha_{+}=\beta_{-}=0 .
\end{equation*}
The up spin is now found in the downward direction whereas the down spin appears only in the upward direction, as is clear from the solution presented in eq. \ref{eq:soln}. This goes absolutely against the dictates of conventional wisdom. It is evident that the angle $\theta$ plays an important role in determining the final state of the particle which leaves the apparatus and evolves freely  to the screen. 

Furthermore, in the limit of a strong magnetic field, we can always get back the well known text book result of SG experiment where the Larmor frequency $\omega_{0}$ is much larger than $\omega$. Here,
\begin{eqnarray} 
\alpha_{+} = e^{-i\frac{\omega_{0}t}{2}};\qquad \alpha_{-} = 0\nonumber\\
\beta_{-} = e^{i\frac{\omega_{0}t}{2}} ; \qquad \beta_{+} = 0~, \label{largeomega}
\end{eqnarray}
as
\begin{equation}
\bar{\omega}= \omega_{0}\sqrt{1+\frac{\omega^{2}}{\omega_{0}}-2\frac{\omega \cos\theta}{\omega_{0}}}\approx \omega_{0}\label{large}.
\end{equation}
Now, plugging this in equation \ref{eq:soln}, we have
\begin{equation}
\Psi(t)
= c_{+}e^{-i\frac{\omega_{0}t}{2}}|+\rangle_{u_0}\otimes \phi_{+}^{\prime}({\bf x^{\prime}},t) 
 +c_{-} e^{i\frac{\omega_{0}t}{2}}|-\rangle_{u_0}\otimes\phi_{-}^{\prime}({\bf x^{\prime}},t) \label{last}
\end{equation}

This is equivalent to the solution \ref{ent}  obtained in references \cite{platt1992,oliveira2006} for a time averaged magnetic field with fixed  direction. One can easily realize that in this case the up (down) spin is found only  in the upward (downward) direction. Thus we arrive at the conventional result of path spin entanglement from our analysis in a limiting situation where the adiabatic evolution is permissible. Note that we have not introduced any averaging concept in Schr\"odinger equation thereby ensuring that the solution holds good at arbitrary time scales. It is clear from eq. \ref{last} that the final state if observed above a time scale $t=\frac{2\pi}{\omega_{0}}$ reduces to a solution of Pauli equation given in \cite{platt1992}


\section{Discussion and Conclusion}

In this article, we have investigated the interplay between Larmor precession and azimuthal inhomogeneity of magnetic field in the context of path-spin entanglement of a spin half system in an SG experiment. Emphasis has been laid on determining the necessary and sufficient condition for adiabatic evolution of spin states owing to its importance in spin measurement using SG apparatus. To the best of our knowledge, the phenomenon predicted by us has not been explored in extant literature on this topic. Earlier studies on dynamical nature of entanglement also took into consideration the effect of inhomogeneous magnetic fields, a relevant example being the article by Caldeira et al. \cite{oliveira2006}.
But the authors in \cite{oliveira2006} presented the analysis with the help of a magnetic field which is changing  in magnitude only. It is obvious that  the instantaneous states will not then change with time. A similar setup may also be incorporated in our work to get back the result of  \cite{oliveira2006}. For the sake of simplicity and clarity, we have restricted ourselves to an azimuthally inhomogeneous magnetic field and postponed a study of the more general situation to a future article  \cite{nirupamprep2014ent}. Our present analysis reveals that not only the ratio of $\dot{\Phi}(t)$ to $\omega_{0}$ but the angle $\theta$ is also a key factor in determining the category of evolution of spin states. Setting $\theta = 0$ takes us back to the standard text book result for SG experiment even if we choose a description in terms of instantaneous states. When $\theta$ is not sufficiently small, unless $\omega_0 >> \dot{\Phi}$, the spin states  evolve non-adiabatically. Though entanglement can still persist, it will no longer allow one to infer any definite spin from the trajectory of a particle. It must be carefully noted that the observation of both spins in either direction, as discussed here, is completely independent of formal and operational non-idealness.  Rather, we stress that the interplay between inhomogeneity of magnetic field and  Larmor precession is quite generic and can lead to novel unexpected phenomena.

There is a practical applicability of this finding, as well. Till date, an ideal SG apparatus has often been used as a beam splitter in quantum information devices.  We have seen in section $3$ that due to a non-adiabatic evolution, a part (some times all) of the up (down) spin can appear in the downward (upward) direction,  keeping the total spin of the system conserved. This can provide a spin flipping tools in various quantum information protocols.

\ack
Our sincere thanks are due to Nicolas Borghini and Debashis Mukherjee for their insightful advice. We thank Archan. S. Majumder for suggestin reference \cite{home2007} which helped us to understand operational and formal non-idealness in SG apparatus.  N.D acknowledges support from DAE, Govt. of India and A. D. acknowledges DST, Govt. of India  for supporting the research.

\section*{References}
\bibliographystyle{iopart-num}
\bibliography{entanglement}{}

\end{document}